\documentclass[aps,prd,eqsecnum,nofootinbib]{revtex4}

\usepackage{graphicx}
\usepackage{hyperref}
\usepackage{amsmath,amsfonts,amssymb,color,latexsym,theorem,mathrsfs,comment,color}

\newcommand{\ma}[1]{\mbox{$\mathcal{#1}$}}

\newcommand{\D}{{\rm d}}

\newcommand{\ti}{\tilde}
\newcommand{\we}{\wedge}

\begin{document}

\title{Ellis-Bronnikov wormholes with NUT charge}
\author{Masato Nozawa,${}^1$ Tomohiro Harada,${}^2$ and Keisuke Nakashi${}^{2,3}$}

\address{
${}^1$Department of General Education, Faculty of Engineering, 
Osaka Institute of Technology, 5-16-1, Omiya, Asahi-ku, 
Osaka, Osaka 535-8585, Japan \\
${}^2$Department of Physics, Rikkyo University, Toshima, Tokyo 171-8501, Japan \\
${}^3$Kaichi Tokorozawa Secondary School, 169 Ohaza Matsugo,
Tokorozawa City, Saitama 359-0027, 
Japan
}


\begin{abstract} 
We construct an exact NUT-charged generalization of the Ellis-Bronnikov wormhole in the Einstein-phantom-scalar system. 
The solution is stationary, asymptotically locally flat at both ends and free of curvature singularities. 
It inherits characteristic features of the Taub-NUT geometry, including the Dirac-Misner string and closed timelike curves, while the absence of Killing horizons eliminates the horizon-induced infinite winding of geodesics and renders the spacetime geodesically complete. 
By allowing the phantom field to have a nontrivial scalar potential with an infinite sequence of AdS vacua, 
we also obtain an exact asymptotically locally AdS wormhole with a NUT charge, which interpolates between different vacua at its two asymptotic ends. 
Using the covariant phase space formalism, we show that the Hamiltonian charges associated with the two asymptotic regions are finite under appropriate scalar boundary conditions. 
By performing a double Wick rotation to Euclidean signature, 
we demonstrate that the conformal class of the resulting metric admits an ambi-K\"ahler structure.
\end{abstract}

\maketitle

\section{Introduction}

The vacuum solution to Einstein's equations discovered by Taub \cite{Taub:1950ez}, 
and Newman, Tamburino and Unti (NUT) \cite{Newman:1963yy} may be regarded as the gravitomagnetic counterpart of the Schwarzschild solution, with the NUT charge playing the role dual to the ordinary mass~\cite{Dowker:1974znr,Argurio:2008zt,Galtsov:2026qlf}. The Taub-NUT metric can be written in a stationary form as  
\begin{align}
\label{TaubNUT}
\D s^2=-\frac{r^2-2mr-n^2}{r^2+n^2}(\D t+2n \cos\theta \D \phi)^2
+(r^2+n^2)\left(\frac{\D r^2}{r^2-2m r-n^2}+\D \theta^2+\sin^2\theta \D \phi^2 \right)\,,
\end{align}
where $m$ is the mass and $n$ is the NUT parameter. 
Despite the apparent  simplicity of this metric form, the Taub-NUT spacetime is notoriously fraught with pathological features that render it physically problematic as a realistic astrophysical model~\cite{Misner:1963fr}. Specifically, the presence of a non-zero NUT charge introduces a Dirac-Misner string, which is a coordinate/gauge-like string singularity associated with the nontrivial fibration of the time direction over the two-sphere. Removing the Dirac-Misner string globally by identifying the time coordinate periodically comes at the price of introducing closed timelike curves (CTCs), thereby violating causality. The periodic identification of the time coordinate also affects the extension of the metric across the Killing horizons: a Hausdorff analytic extension is geodesically incomplete, whereas geodesic completeness can be achieved only at the price of allowing a non-Hausdorff extension \cite{Miller:1971em}. 
Furthermore, the spacetime is only asymptotically locally flat (ALF), even though the Riemann tensor tends to zero at large radius. 
Because of the accumulation of these anomalies, Misner famously described the Taub-NUT geometry as ``a counterexample to almost anything'' \cite{Misner:1965zz}.

Despite these unconventional global properties, the Taub-NUT spacetime possesses 
a remarkably rich mathematical structure. A crucial redeeming quality that makes 
the spacetime particularly valuable as a theoretical laboratory is its high degree of mathematical tractability. The spacetime is algebraically special, belonging to Petrov type D, and admits an enhanced set of spacetime symmetries. Moreover, 
the geodesic equations and various test-field equations are completely separable, 
allowing its gravitational dynamics to be analyzed in considerable detail.

The nontrivial asymptotic structure also makes the Taub-NUT geometry a useful 
testing ground for gravitational magnetic charges and their associated conserved 
quantities. This aspect is closely related to formulations of spatial infinity, 
such as the Ashtekar-Hansen framework \cite{Ashtekar:1978zz}, in which the 
electric and magnetic parts of the asymptotic Weyl tensor play a central role.

A different class of geometries in which global properties of spacetime play an equally prominent role is provided by traversable wormholes.
Such geometries connect two otherwise distinct regions of spacetime through a throat and have 
long served as theoretical laboratories for exploring the interplay between spacetime topology, causality, and energy conditions.
Among the simplest examples is the static solution supported by a massless phantom scalar field, independently obtained by Ellis \cite{Ellis:1973yv} and Bronnikov \cite{Bronnikov:1973fh} in 1973. The underlying physical conditions for a traversable wormhole, and their implications for violations of the energy conditions, were later systematically formulated by Morris and Thorne in 1988  \cite{Morris:1988cz, Morris:1988tu}.
Although traversable wormholes have long been an arena for paradoxes like CTCs, they have attracted renewed
attention in recent years, particularly in connection with 
quantum entanglement and traversability in gauge/gravity duality  \cite{Maldacena:2004rf,Gao:2016bin}.

Traversable wormholes including the Ellis-Bronnikov one have also been intensively studied in the context of astrophysics 
as one of the black hole mimickers. Wormholes are particularly interesting black-hole mimickers because, like black holes, 
they possess no material surface. 
Since astrophysical observations of black-hole candidates are usually carried out by photons, understanding the structure of null geodesics is of 
key importance. In this context, it should be noted that the Ellis-Bronnikov wormholes have photon spheres as black holes. They involve particular features in gravitational lensing, which may make them observationally distinguishable from 
black holes, see e.g.,~\cite{Chetouani:1984qdm,Dey:2008kn,Abe:2010ap,Nakajima:2012pu,Tsukamoto:2012xs,Yoo:2013cia,Tsukamoto:2016zdu,Aounallah:2020rlf,Ohgami:2015nra,Takahashi:2026hfp}.

A natural question then arises: what becomes of the global and causal structure 
of an Ellis-Bronnikov wormhole when a gravitomagnetic NUT charge is switched on?
Wormholes carrying a NUT charge have previously been investigated in several 
different settings, including self-gravitating Skyrmion systems \cite{Ayon-Beato:2015eca}, nonlinear electrodynamics 
\cite{Flores-Alfonso:2020nnd}, and modified-gravity theories \cite{Ibadov:2020ajr,Ibadov:2021oqf}. 
It is therefore natural to ask whether the 
simplest Ellis-Bronnikov system itself admits an exact NUT-charged generalization.
A particularly close analogue is provided by the overcharged  Reissner-Nordstr\"om-Taub-NUT solution,
which describes a two-ended NUT wormhole in Einstein-Maxwell theory~\cite{Clement:2015aka}. 

In this paper, we construct an exact NUT-charged generalization of the Ellis-Bronnikov wormhole and explore its physical properties. 
Exploiting the sigma-model symmetries of the stationary Einstein-phantom-scalar system,  
we generate a new solution by means of an Ehlers-type transformation. 
The resulting spacetime is ALF at both ends and is  free of curvature singularities. A notable distinction from the ordinary Taub-NUT spacetime 
is the absence of Killing horizons, which allows one to avoid the infinite winding of geodesics around the horizon. 
On the other hand, the solution shares some characteristic global features with the Taub-NUT geometry, 
including the presence of a Dirac-Misner string and CTCs.

We further construct the asymptotically locally anti-de Sitter (AdS) solution, by introducing the 
potential for the phantom scalar field. The potential admits an infinite number of AdS vacua and 
is obtained  through  a suitable analytic continuation of ${\cal N}=2$ gauged supergravity. 
As in the vanishing NUT case, the phantom scalar approaches the origin of field space at one asymptotic end, 
whereas at the other end it approaches a neighboring AdS vacuum, whose location depends on the branch of the solution. 
We determine the mass associated with each asymptotic region using the covariant phase space formalism. 
We also present an ALF Euclidean counterpart of the solution, which may be useful in the study of gravitational instantons. 

The organization of the present article is as follows. In the next section, we derive a new nutty wormhole solution
by a solution-generating technique. Various physical properties of the solution are addressed in section \ref{sec:phys}. 
Section~\ref{sec:AdS} explores the asymptotically locally AdS solutions.
Our conclusions are summarized in section~\ref{sec:conc}. 
An appendix~\ref{sec:Euc} investigates the Euclidean counterpart.

\section{Asymptotically locally flat solutions}
\label{sec:asyflat}

We consider a phantom scalar field in 4-dimensional general relativity, for which 
the Lagrangian is given by 
\begin{align}
\label{}
\ma L= R+  \frac 12 (\nabla \varphi)^2 \,. 
\end{align}
Here, the kinetic term of the scalar field has the opposite sign. 
Apart from adopting the units $8\pi G=c=1$, we use the metric signature $(-,+,+,+)$ and 
employ the curvature conventions of~\cite{Wald:1984rg} throughout this paper.
Einstein's equations and the scalar field equations then read
\begin{align}
\label{Eineq4D}
R_{\mu\nu}= -\frac 12 \nabla_\mu \varphi \nabla_\nu \varphi \,, \qquad \nabla^2 \varphi=0\,. 
\end{align}

Although our primary aim is to construct a NUT-charged wormhole, solving the nonlinear system (\ref{Eineq4D}) 
directly is rather cumbersome. Instead, it is considerably simpler to employ a solution-generating method.

\subsection{Solution-generating method via coset space symmetry}

We are interested in the stationary solutions in this theory, for which the local metric and the 
scalar field take the $t$-independent form 
\begin{align}
\label{stationarymetric}
\D s^2=-f \left(\D t+\varpi_i \D x^i\right)^2 +f^{-1} h_{ij} \D x^i \D x^j \,, \qquad 
\varphi=\varphi(x^i) \,,
\end{align}
where $h_{ij}$ is the 3-dimensional metric of the orbit space of the 
stationary Killing vector $\xi=\partial/\partial t$. $f$ and $\varpi$ are the scalar and  a 1-form on the orbit space, 
corresponding to the norm and the rotation of $\xi$. 
Denoting the twist 1-form as 
$\omega_\mu \equiv \epsilon_{\mu\nu\rho\sigma}\xi^\nu \nabla^\rho \xi^\sigma$, 
a standard computation shows that $\omega $ is closed 
$\nabla_{[\mu }\omega_{\nu]} =0$ (see (7.1.15) of \cite{Wald:1984rg}), provided that Einstein's equations (\ref{Eineq4D}) hold. 
Then, there exists a local scalar $\psi$ such that 
$\omega_\mu =\nabla_\mu \psi$ with $\xi^\mu \nabla_\mu \psi=0$. 
In terms of the metric components (\ref{stationarymetric}), the twist 1-form 
$\omega_\mu = \epsilon_{\mu\nu\rho\sigma}\xi^\nu \nabla^\rho \xi^\sigma$ translates into
\begin{align}
\label{varpieq}
\D \psi=f^2 \star_3 \D \varpi \,,
\end{align}
where $\star_3$ denotes the Hodge dual for $h_{ij}$.

Following the formulation in \cite{Galtsov:1995mb}, field equations are then decomposed into 
\begin{align}
\label{3DEineq}
{}^{(3)\!}R_{ij}= \frac 1{2f^2}(D_i fD_jf+D_i\psi D_j \psi)-\frac 12 D_i\varphi
D_j \varphi \,,
\end{align}
and 
\begin{align}
\label{}
fD^iD_i f=(Df)^2 -(D \psi)^2 \,, 
\qquad 
D^i (f^{-2}D_i \psi)=0 \,, 
\qquad D^iD_i\varphi=0\,, 
\end{align}
where $D_i$ and ${}^{(3)\!}R$ denote the covariant derivative and Ricci scalar 
for $h_{ij}$, respectively. 
These equations are derived from the following 3-dimensional Lagrangian 
\begin{align}
\label{}
\ma L_3 = {}^{(3)\!}R -h^{ij}G_{AB}(\phi)D_i \phi^A D_j \phi^B\,, \qquad 
\phi^A=(f, \psi, \varphi)\,,
\end{align}
where the target space metric $\D s_T^2=G_{AB}(\phi)\D \phi^ A \D \phi^B$ is described by  
$\mathbb R\times {\rm SL}(2, \mathbb R)/{\rm SO}(2)$ coset space \cite{Galtsov:1995mb}
\begin{align}
\label{}
\D s_T^2 =\frac 1{2f^2}(\D f^2+\D \psi^2)-\frac 12 \D \varphi^2 \,. 
\end{align}

We impose the boundary condition of ``local asymptotic flatness''\footnote{
In this paper, we define a stationary spacetime to be ALF if it satisfies the boundary conditions in (\ref{BC}). 
Compared with the standard notion of asymptotic flatness, our definition is more restrictive in requiring stationarity, 
but less restrictive in allowing a nonvanishing NUT charge. 
In the Ashtekar-Hansen framework~\cite{Ashtekar:1978zz}, the asymptotic symmetry group at spatial infinity is the {\it Spi} group. 
Imposing the additional condition that the leading magnetic part of the Weyl tensor vanishes selects a preferred Poincar\'e subgroup, 
and excludes spacetimes carrying NUT charge. 
We also note that the coefficient of the $1/r$ term of $\psi$ is proportional to the NUT charge. In the absence of the NUT charge, as in the Kerr spacetime, the asymptotic behavior is $\psi = O(1/r^2)$.}
\begin{align}
\label{BC}
f=1+O(1/r)\,, \qquad \psi=O(1/r) \,, \qquad h_{ij}=\delta_{ij}+O(1/r) \,,
\qquad \varphi=\bar\varphi_\infty+O(1/r)\,, 
\end{align}
where  $\bar\varphi_\infty$ is the asymptotic value of the scalar field and 
$r=(\delta_{ij}x^i x^j)^{1/2}$ is the radial coordinate associated with the asymptotically Cartesian coordinates $x^i$. 
The Taub-NUT spacetime (\ref{TaubNUT}) is indeed subjected to these boundary conditions.

We now focus on the 2-dimensional part 
$\D s_{T (2)}^2=\frac 1{2}f^{-2}(\D f^2+\D \psi^2)$ of the target space. 
Let us define the following $2\times 2 $ real symmetric matrix 
\begin{align}
\label{}
\ma M=\left(
\begin{array}{cc}
 f+f^{-1}\psi^2     &  -f^{-1}\psi   \\
-f^{-1}\psi      &   f^{-1} 
\end{array}
\right)\,,\qquad {\rm det}\ma M=1 \,, 
\end{align}
in terms of which 
\begin{align}
\label{}
\D s_{T (2)}^2=-\frac 14 {\rm Tr}(\D \ma M \D (\ma M^{-1})) \,. 
\end{align}
This metric is invariant under the global Ehlers ${\rm SO}(2)$ transformation~\cite{Stephani:2003tm}
\begin{align}
\label{Omat}
\ma M \to \ma M'=O\ma M O^T \,, \qquad 
O=\left(
\begin{array}{cc}
 \cos a     &  -\sin a   \\
\sin a       &   \cos a 
\end{array}
\right)\,,
\end{align}
where $a$ is a constant. It should be emphasized that this transformation preserves the 
ALF boundary conditions (\ref{BC}). This means that one can generate 
a new ALF solution from an old one. 
Specifically, if the seed solution is static 
$\phi^A_0=(f_0, \psi_0=0, \varphi_0)$, the generated solution takes the form 
\begin{align}
\label{trans}
f=&\,\frac{f_0}{\cos^2 a+f^2_0\sin^2 a } \,, \qquad 
\psi=\frac{(1-f^2_0)\sin a \cos a}{\cos^2 a+f^2_0\sin^2 a }\,, \qquad 
\varphi=\varphi_0 \,. 
\end{align}

When the Schwarzschild metric ($f_0=1-r_0/r$, $\varphi_0=0$) is employed as a seed, 
this transformation allows us to obtain the Taub-NUT metric (\ref{TaubNUT})
in a purely algebraic manner, see e.g., \cite{Stephani:2003tm}.

\subsection{Background solution}

As a background seed solution, let us consider the static 
Ellis-Bronnikov wormhole solution \cite{Ellis:1973yv,Bronnikov:1973fh}
\begin{align}
\label{EBmetric}
f_0 =e^{-2\beta U} \,, \qquad 
\varphi_0=\pm 2\sqrt{1+\beta^2} U \,, \qquad 
h_{ij}\D x^i \D x^j =\D r^2+(r^2+m_0^2 ) (\D \theta^2+\sin^2\theta \D \phi^2) \,. 
\end{align}
where $\beta $ and $m_0$ are constants, and 
\begin{align}
\label{U}
U(r)\equiv  \frac \pi 2-\arctan \left(\frac{r}{m_0}\right)\,.
\end{align}
This solution is static and spherically symmetric. 
One can also verify that the solution is free of any curvature singularities and is indeed geodesically complete \cite{Ellis:1973yv}. 
The spacetime approaches the Minkowski spacetime at both asymptotic ends $r\to \pm \infty$ with 
Arnowitt-Deser-Misner (ADM) masses~\cite{Arnowitt:1962hi}
\begin{align}
\label{EB:Mass}
M_+= m_0  \beta \,, \qquad M_- = -  m_0 \beta  e^{\pi \beta}\,. 
\end{align}

Since the sign flip $m_0\to -m_0$ can be compensated by $r\to -r$, 
one can assume $m_0>0$ without losing any generality. 
Furthermore, the transformation $\beta\to-\beta$, together with an interchange of the two asymptotic regions 
and the redefinition $m_0\to e^{\pi\beta}m_0$, leaves the geometry unchanged. 
We may therefore restrict attention to $\beta\ge 0$.
The special case $\beta=0$ corresponds to the ultrastatic massless wormhole, 
for which the geometry admits the reflection symmetry $r\to -r$. 
In contrast, any nonzero value of $\beta $ breaks this reflection symmetry and yields inequivalent asymptotic regions with different ADM masses.

For further details of the Ellis-Bronnikov wormholes, 
we refer the reader e.g.,  to \cite{Martinez:2020hjm,Nozawa:2020wet} for global structures, 
to \cite{Gonzalez:2008wd,Torii:2013xba,Cremona:2018wkj} for linear instability, 
to \cite{Chetouani:1984qdm,Dey:2008kn,Abe:2010ap,Nakajima:2012pu,Tsukamoto:2012xs,Yoo:2013cia,Tsukamoto:2016zdu,Aounallah:2020rlf,Takahashi:2026hfp} for gravitational lensing and black-hole shadows.

\subsection{Ellis-Bronnikov wormhole with NUT charge}

We are now in a position to obtain the desired wormhole with a NUT charge. 
We employ the Ellis-Bronnikov metric (\ref{EBmetric}) as a seed and 
rewrite the rotation angle $a$ as 
$\cos (2a)=m/m_0$, $\sin (2a)=n/m_0$, i.e., 
\begin{align}
\label{constraint}
m_0=\sqrt{m^2+n^2}\,. 
\end{align}
Applying the above transformation (\ref{trans}), we obtain
\begin{align}
\label{fpsi}
f(r)=&\,\frac{m_0}{m_0 \cosh (2\beta U(r))+m \sinh (2\beta U(r))}\,, \\
\psi(r)=&\,\frac{n\sinh (2\beta U(r))}{m_0 \cosh (2\beta U(r))+m \sinh (2\beta U(r))}\,. 
\end{align}
From  (\ref{varpieq}), we get $\varpi=2n\beta \cos\theta \D \phi$. 
We therefore arrive at 
\begin{align}
\label{metric}
\D s^2=&\,-f (r) (\D t+2n\beta \cos \theta \D \phi )^2 
+\frac 1{f(r)}[\D r^2+(r^2+m_0^2)(\D \theta ^2 +\sin^2 \theta \D \phi^2)]\,, 
\\
\label{scalar}
\varphi=&\, \pm 2\sqrt{1+\beta ^2 }U(r)\,. 
\end{align}
This spacetime can be identified as the Ellis-Bronnikov wormhole with a NUT charge. 
To the best of our knowledge, this solution appears to be original.

\section{Physical properties}
\label{sec:phys}

This section explores various physical properties of the stationary solution (\ref{metric}), 
which is characterized by 3 parameters $m, \beta $ and $n$. 
The first two quantities already appear in the Ellis-Bronnikov solution (\ref{EBmetric}), characterizing 
the  mass and scalar charge, while $n$ corresponds to the NUT parameter (see (\ref{MNphi+}) below). 
Since the solution (\ref{metric}) is invariant under $\beta \to -\beta$, $m\to -m$ and $n\to -n$, 
we can set $\beta \ge 0$. Moreover, $n\to -n$ can be compensated by changing the direction of rotation $\phi\to -\phi$. 
It follows that 
we can restrict ourselves, without loss of generality,  to
\begin{align}
\label{}
\beta \ge 0 \,, \qquad n \ge 0 \,. 
\end{align}

The  first important observation is that 
 $f(r)$ in (\ref{fpsi}) satisfies 
 \begin{align}
\label{}
f(r)>0 \,, 
\end{align}
throughout the entire range $r\in \mathbb R$. 
Consequently, the spacetime contains no Killing horizons. 
This property distinguishes the present solution from the conventional Taub-NUT spacetime (\ref{TaubNUT}), 
which admits two distinct Killing horizons.

\subsection{Curvature invariants}

The scalar curvature $R$ and the square of the Weyl tensor $\ma W^2\equiv C_{\mu\nu\rho\sigma}C^{\mu\nu\rho\sigma}$ read 
\begin{align}
\label{}
R=&\, f''-\frac{3f'{}^2}{2f}+\frac{2rf'}{r^2+m_0^2}+\frac{2f(n^2\beta^2 f^2-m_0^2)}{(r^2+m_0^2)^2}\,,
\\
\ma W^2=&\,\frac{4}{3(r^2+m_0^2)^4}[(r^2+m_0^2)\{(r^2+m_0^2)f''-rf'\}-2f(m_0^2+2n^2\beta^2 f^2)]^2 
\notag \\ &\, 
-\frac{48 n^2\beta^2 f^2}{(r^2+m_0^2)^4}[(r^2+m_0^2)f'-rf]^2 \,. 
\end{align}
We can see that these quantities remain finite for all $r\in \mathbb R$. 
Although the Kretschmann scalar $R_{\mu\nu\rho\sigma}R^{\mu\nu\rho\sigma}$ does not have an illuminating expression, 
it exhibits a similar qualitative behavior and is likewise finite everywhere. 
This is a typical feature of wormhole geometries.

It is expedient to spell out the algebraic structure of the Weyl tensor. 
A direct computation shows that the spacetime is of Petrov type D. 
Thus, despite the presence of the phantom scalar field and the
nontrivial wormhole geometry, the solution belongs to the same algebraically 
special class as for the ordinary Taub-NUT spacetime (\ref{TaubNUT}).

It should be noted, however, that the above analysis only excludes curvature singularities represented by scalar invariants. 
As we discuss in section~\ref{sec:DiracMisner}, it does not rule out the presence of distributional singularities supported on lower-dimensional submanifolds.

\subsection{Infinity and conserved charges}

In the asymptotic limit $r\to \infty$, we see 
\begin{align}
\label{asymex1}
f=1-\frac{2m\beta}{r} +O(1/r^2)\,, \qquad 
\psi=\frac{2n \beta }{r}+O(1/r^2) \,,\qquad 
\varphi=\pm \frac{2\sqrt{(1+\beta^2)(m^2+n^2)}}{r}+O(1/r^2)\,.
\end{align}
Thus, the spacetime is ALF for $r\to \infty$
in the sense of obeying the boundary conditions (\ref{BC}). 
Since the metric is stationary with the Killing vector $\xi=\partial/\partial t$, 
the mass is given by a Komar charge~\cite{Komar:1958wp} (see also \cite{Manko:2005nm,Galtsov:2026wxl}), 
whereas
the NUT charge is defined by the gravitational dual of the Komar charge as \cite{Bossard:2008sw,Bossard:2009at,Godazgar:2018qpq}
\begin{align}
\label{}
 M=-\frac 1{8\pi} \int _{S^2} \star \D \xi \,, \qquad 
 N= \frac 1{8\pi} \int _{S^2} \D \xi \,.
\end{align}
Here the integration is performed over a 2-sphere at infinity, and 
we use the same symbol $\xi$ for the metric dual one-form of the Killing vector to lighten the notation.
The scalar charge is given by
\begin{align}
\label{}
\varphi^{(\infty)} =-\frac 1{4\pi} \int_{S^2} \star (\xi \we \D \varphi ) \,. 
\end{align}
Inserting (\ref{asymex1}), we obtain the conserved quantities in the $r>0$ region as 
\begin{align}
\label{MNphi+}
M_+=m \beta \,, \qquad N_+=n \beta \,, \qquad 
\varphi_+^{(\infty)} =\pm 2\sqrt{(1+\beta^2)(m^2+n^2)} \,. 
\end{align}
Note that one cannot take the $\beta \to 0$ limit while the NUT charge $N$ nonvanishing. 
Hence, we shall confine ourselves to the $\beta >0$ case hereafter.  
The zero mass solution $M_+=0$ is alternatively obtained by $m=0$. 

Let us next investigate the behavior in the domain of large negative $r$. 
With this objective in mind, we introduce rescaled coordinates
\begin{align}
\label{titr}
\ti t=\kappa^{1/2} t \,, \qquad 
\ti r=-\kappa^{-1/2} r \,,
\end{align}
where 
\begin{align}
\label{kappa}
\kappa  \equiv \frac{m_0}{\bar m_0}\,,
\end{align}
and
\begin{align}
\label{}
\bar m_0\equiv m_0\cosh (2\pi \beta)+m \sinh(2\pi \beta )\,. 
\end{align}
In the limit $\ti r\to \infty$, we have $f\simeq \kappa(1-2M_-/\ti r)+O(1/\ti r^2)$, where 
\begin{align}
\label{Mminus}
M_- = - \kappa^{1/2} \bar m \beta\,,
\end{align}
and
\begin{align}
\label{}
\bar m \equiv m \cosh (2\pi \beta)+m_0 \sinh(2\pi \beta )\,. 
\end{align}
Here we notice the useful relation $\bar m_0^2-\bar m^2=n^2$. 
We thus find 
\begin{align}
\label{}
\D s^2\simeq &\,-\left(1-\frac{2M_-}{\ti r}\right)\left(\D \ti t+2\ti n\beta\cos\theta \D \phi \right)^2 +\left(1+\frac{2M_-}{\ti r}\right)
\left[\D \ti r^2+\ti r^2(\D \theta ^2+\sin^2\theta \D \phi^2)\right]\,, 
\\
\varphi\simeq &\, \pm \left(2\pi \sqrt{1+\beta^2}-\frac{2 \sqrt{1+\beta^2}m_0}{\kappa^{1/2}\ti r}\right) \,. 
\end{align}
where
\begin{align}
\label{tin}
\ti n\equiv n \kappa^{1/2} \,. 
\end{align}
Thus, $M_-$ represents the ADM mass in the $\ti r>0$ region. 
It should be noted here that the time translation Killing vector in the $r<0$ region is normalized 
as $\partial/\partial \ti t$. 
Similarly,  the NUT and the scalar charges in the $r<0$ region read
\begin{align}
\label{}
N_-=&\, \ti n\beta  \,, \\
\varphi_-^{(\infty)}=&\, \mp \frac{2\sqrt{1+\beta^2}m_0}{\kappa^{1/2}} \,. 
\end{align}
In the limit of vanishing NUT parameter $n\to 0$, 
$M_-$ in (\ref{Mminus})  recovers (\ref{EB:Mass}). 
These conserved quantities fulfill the following relations
\begin{align}
\label{}
M_+^2+N_+^2 = m_0^2 \beta^2\,, \qquad 
M_-^2+N_-^2=\bar m_0^2 \beta^2 \kappa \,. 
\end{align}

\subsection{Energy conditions and throat}
\label{sec:energycond}

It has been well recognized that the wormhole throat structure is realized, at least in the static and spherically symmetric case,  only when the 
null energy condition is violated \cite{Morris:1988cz,Morris:1988tu}. In other words, there must exist ``exotic matter'' at the throat. 
In the present setting, this is supplied by the phantom scalar field, whose stress-energy tensor
is given by
\begin{align}
\label{}
T_{\mu\nu}=-\frac 12 \left(\nabla_\mu\varphi\nabla_\nu\varphi-\frac 12 (\nabla\varphi)^2g_{\mu\nu}\right)\,.
\end{align}
The $tt$-component of the stress-energy tensor of the phantom scalar field  is reduced to 
\begin{align}
\label{}
T_{tt}=-\frac{m_0^2(1+\beta^2)f(r)^2}{(r^2+m_0^2)^2}<0 \,. 
\end{align}
It follows that the weak energy condition \cite{Wald:1984rg} is violated everywhere in the spacetime. 

In the spherically symmetric case, the locus of the wormhole throat is geometrically 
identified in terms of the areal radius of round $S^2$ without ambiguity.  
In the presence of NUT charge, however, the situation is more subtle. Although the spacetime admits an ${\rm SU}(2)$ isometry, 
its group orbits do not define a family of globally well-defined spacelike round two-spheres. 
Indeed, the NUT charge induces a nontrivial fibration of the time direction over $S^2$, encoded in the one-form (\ref{oneformA}). 
Consequently, there is no invariant areal radius defined directly from the area of round $S^2$ surfaces in the full spacetime, 
and the location of a wormhole throat cannot be characterized in the usual spherically symmetric manner.

Following~\cite{Ayon-Beato:2015eca,Clement:2015aka},  it is still reasonable to define the following areal radius  of the $S^2$ base in the three-dimensional orbit space
 \begin{align}
\label{}
S(r) \equiv g_{\theta\theta}^{1/2} =\sqrt{\frac{r^2+m^2_0}{f(r)}}\,. 
\end{align}
The minimum of $S(r)$ defines a minimal two-sphere, 
and hence a throat-like bottleneck, in the three-dimensional orbit space of the stationary Killing field.
This orbit-space interpretation has a natural physical as well as geometrical meaning, since dimensional reduction along the stationary Killing field yields a three-dimensional theory of gravity coupled to matter fields, as in (\ref{3DEineq}). It therefore provides a more appropriate framework for the present stationary geometry than an ADM decomposition based on constant-time slices. From this viewpoint, the usual throat definition in the static case may also be reinterpreted in terms of the Killing orbit space rather than specifically in terms of a constant-time slice.
In what follows, we refer to such a minimal two-sphere as the ``throat.''

The locus of the throat is determined implicitly by the root of the following equation
\begin{align}
\label{throateq}
e^{4\beta U(r)}(m_0+m)(r-m_0 \beta)+(m_0-m)(r+m_0\beta)=0 \,. 
\end{align}
The throat equation (\ref{throateq}) admits a unique real root $r=r_{\rm th}$. 
To see this concretely, let us introduce $x=r/m_0$ and $\Gamma=m/m_0$. 
Due to $|\Gamma|<1$, any root must lie in the interval $-\beta<x<\beta$.
Within this interval, the throat equation (\ref{throateq}) is equivalent to
\begin{align}
H(x)
\equiv
\ln\left(\frac{x+\beta}{\beta-x}\right)
-4\beta\left(\frac{\pi}{2}-\arctan x\right)
-\ln\left(\frac{1+\Gamma}{1-\Gamma}\right)
=0.
\end{align}
Since $H'(x)=2\beta /(\beta^2-x^2)+4\beta /(1+x^2)>0$, 
it turns out that $H(x)$ is monotonically increasing. 
Moreover, 
\begin{align}
\lim_{x\to-\beta+}H(x)=-\infty\,, 
\qquad
\lim_{x\to\beta-}H(x)=+\infty\,, 
\end{align}
leading to the existence of a root by the intermediate value theorem. 
It follows from monotonicity of $H(x)$ that the throat equation possesses exactly one real root $r=r_{\rm th}$, 
which satisfies
\begin{align}
-m_0\beta<r_{\rm th}<m_0\beta\,.
\end{align}

\subsection{CTCs \& Dirac-Misner string}
\label{sec:DiracMisner}

The present metric (\ref{metric}) includes the following off-diagonal term
\begin{align}
\label{oneformA}
\varpi= 2n\beta \cos \theta \D \phi \,. 
\end{align}
This can be viewed as a 1-form on $S^2$. 
Writing $F=\D \varpi$, we can define the associated magnetic charge 
\begin{align}
\label{NUTchern}
N \equiv - \frac 1{8\pi}\int_{S^2}F  \,,
\end{align}
giving rise to $N_+=n \beta$. 
If the 1-form $\varpi$ were globally defined on $S^2$, Stokes' theorem implies 
$\int _{S^2} \D \varpi=\int_{\partial S^2}\varpi=0$, in contradiction to the nonvanishing flux above. 
Thus, the 1-form is defined only locally. Equivalently, the curvature two-form $F$ represents a nontrivial cohomology class 
in the second de Rham cohomology group  $H^2(S^2)$, representing the space of closed 2-forms modulo exact two-forms \cite{Eguchi:1980jx}. 
The dimension of this group is given by the second Betti number as $b_2(S^2)=1$.

The lack of global regularity of the above 1-form can also be seen explicitly by rewriting it in Cartesian coordinates. 
Introducing 
$x+iy =r \sin\theta e^{i\phi}$, $z=r \cos\theta$, we have 
\begin{align}
\label{}
\varpi =\frac{2n \beta z(-y \D x+x \D y)}{(x^2+y^2)\sqrt{x^2+y^2+z^2}}\,,
\end{align}
implying that this 1-form $\varpi$ is singular on the entire $z$-axis $x^2+y^2=0$. 
Thus, the 1-form cannot be represented by a single regular expression as $\varpi=2n\beta\cos\theta \D\phi$ 
covering the whole two-sphere. This singularity is the gravitational analogue of the Dirac string associated with a magnetic monopole 
and is referred to as the Dirac-Misner string. 
This singularity originates from the 1-form ${\rm d}\phi$ itself: while the rotational
Killing vector $\eta=\partial/\partial\phi$ of $S^2$ vanishes on the axis, $\eta^\mu\nabla_\mu\phi=1$ holds everywhere, implying that ${\rm d}\phi$ must be
singular on the axis. 
Hence a 1-form $\alpha(\theta){\rm d}\phi$ is regular on the axis
only if $\alpha=O(\theta^2)$.

To remove the Dirac-Misner string, the time coordinate $t$ must be
periodically identified~\cite{Misner:1963fr}. 
This can be seen as follows. 
Introducing a new time coordinate $t_{N}$ through $t=t_N-2n\beta\phi$ in a patch containing the north pole, we obtain
\begin{align}
\D t+2n\beta\cos\theta\D\phi
&=
\D t_N+2n\beta(\cos\theta-1)\D\phi
\simeq
\D t_N-n\beta\theta^2 \D\phi
\end{align}
near $\theta=0$, and the corresponding one-form $-n\beta\theta^2 \D\phi$ is therefore regular on the positive $z$-axis.
Similarly, in a patch containing the south pole, one may 
introduce another new time coordinate $t_{S}$ through 
$t=t_S+2n\beta\phi$, 
which yields a 1-form regular at $\theta=\pi$. By gluing these two
patches together, the Dirac-Misner string can be removed everywhere.

In the overlapping region of the two patches, the coordinates are related by 
$t_N-t_S=4n\beta\phi$. 
For this gluing to be single-valued, the time coordinate must be periodic. Let $P$ denote the period of $t$, and introduce the $2\pi$-periodic angular coordinate 
$\chi=2\pi t/P$.  The coordinates on the two patches are then related by $\chi_N = \chi_S+8\pi n\beta\phi/P$. 
Consequently, the corresponding transition function is \cite{Eguchi:1980jx}
\begin{align} 
\Psi_{\rm NS}(\phi) = \exp\left(\frac{8\pi i n\beta}{P}\phi\right)\,. 
\end{align} 
This transition function must be single-valued under $\phi\to\phi+2\pi$,
requiring  $\Psi_{\rm NS}(\phi+2\pi)=\Psi_{\rm NS}(\phi) $ or equivalently, $ \exp\left(16\pi^2 i n\beta/P\right)=1$. 
This gives 
 \begin{align} 
 \Delta t=P=\frac{8\pi n\beta}{k}\,, \qquad k\in\mathbb{N}\,. 
 \label{period}
 \end{align}
 This is reminiscent of Dirac's quantization condition.
 
 This quantization condition has a natural interpretation in terms of the topology of the corresponding ${\rm U}(1)$ bundle. 
In terms of the $2\pi$-periodic fibre coordinate $\chi =2\pi t/P$ and (\ref{oneformA}), we define the normalized ${\rm U}(1)$ connection by 
$\mathcal A=2\pi \varpi/P$, whose curvature 2-form is $\mathcal F=\D \mathcal A=2\pi F/P$. 
 The first Chern class is therefore~\cite{Eguchi:1980jx}
 \begin{align}
\label{}
c_1=\left[\frac{\mathcal{F}}{2\pi}\right]
=\left[\frac{F}{P}\right]
\in H^2(S^2,\mathbb{Z})\,, 
\end{align}
The corresponding first Chern number reads 
 \begin{align}
\frac{1}{2\pi}\int_{S^2}\mathcal F
=\frac{1}{P}\int_{S^2}F
=\pm k\,,
\end{align}
where the orientation-dependent sign is understood. 
For $k=1$, the coordinate $\psi=t/(2n\beta)$ has period $4\pi$. A constant-$r$ hypersurface therefore describes a ${\rm U}(1)$ bundle over $S^2$ with first Chern number one and has the topology of $S^3$. For $k>1$, the corresponding constant-$r$ hypersurface has the topology of the lens space $S^3/\mathbb{Z}_k$ \cite{Jante:2013kha,Boulton:2021wln}. 

As discussed above, we must impose a periodic identification on the time coordinate $t$ with period (\ref{period}) to remove the Dirac-Misner string. 
Since the orbits of $\xi=\partial/\partial t$ are closed, they give rise to CTCs whenever $g_{tt}<0$. 
In the present case, we have $f(r)>0$ in the entire spacetime, and hence CTCs are present everywhere.

Alternatively, one may refrain from periodically identifying $t$ and instead tolerate the distributional singularity associated with the Dirac-Misner string, 
following the standpoint first advocated by Bonnor \cite{Bonnor} (see also \cite{Manko:2005nm}). 
In this case, the azimuthal component of the metric is 
\begin{align} 
g_{\phi\phi} = f(r)^{-1}\bigl(r^2+m^2_0\bigr)\sin^2\theta - 4n^2\beta^2 f(r)\cos^2\theta\,. 
\end{align} 
Near either component of the symmetry axis ($\theta=0, \pi $), it behaves as 
\begin{align} 
g_{\phi\phi} \simeq -4n^2\beta^2 f(r)<0\,. 
\end{align} 
Therefore, CTCs are again unavoidable. This behavior is analogous to that found outside the horizon of the ordinary Taub-NUT spacetime.

To conclude this section, the NUT charge cannot be accommodated in a globally regular and causal Lorentzian spacetime within the present construction, irrespective of the presence or absence of the Dirac-Misner string.

\subsection{Geodesics}

We shall now consider the geodesic equation for the metric (\ref{metric}).
For an exhaustive study of geodesic equations in ordinary Taub-NUT spacetime, 
see \cite{Kagramanova:2010bk,Clement:2015cxa}. 

Before going into the details of geodesic equation, it is of use to address the 
spacetime symmetries. We see that the metric admits 4 Killing vectors $\xi_I$ ($I=0,1,2,3$), 
where
$\xi_0\equiv\xi=\partial/\partial t$ is the stationary Killing vector 
\begin{subequations}
\label{KVs}
\begin{align}
\label{}
\xi_0=&\, \frac{\partial}{\partial t}\,, \\
\xi_1=&\,  -\sin\phi \frac{\partial}{\partial \theta}-\cos\phi \cot \theta \frac{\partial}{\partial \phi}
+2n\beta \frac{\cos\phi}{\sin\theta }\frac{\partial}{\partial t}\,, 
\\
\xi_2=&\, \cos\phi \frac{\partial}{\partial \theta}-\sin\phi\cot \theta \frac{\partial}{\partial \phi}
+2n\beta \frac{\sin\phi}{\sin\theta }\frac{\partial}{\partial t}\,, 
\\
\xi_3=&\, \frac{\partial}{\partial \phi}\,, 
\end{align}
\end{subequations}
generating ${\mathfrak u}(1)\oplus {\mathfrak su}(2)$ algebra 
\begin{align}
\label{}
[\xi_0, \xi_i]=0 \,, \qquad [\xi_{i}, \xi_j]=-\epsilon_{ijk} \xi_k \,. 
\end{align}
Here, $\epsilon_{ijk}$ is an alternating tensor with $\epsilon_{123}=1$.

The Hamilton-Jacobi equation takes the form
\begin{align}
\label{}
-\frac{\partial \ma S}{\partial \lambda}=\frac 12 g^{\mu\nu}\frac{\partial \ma S}{\partial x^\mu }\frac{\partial \ma S}{\partial x^\nu }\,,
\end{align}
where $\lambda $ is the affine parameter of the geodesic. 
Consider  the separable form 
\begin{align}
\label{}
\ma S=\frac 12 \mu^2 \lambda - E t +L_{3} \phi +\ma S_r(r)+\ma S_\theta (\theta ) \,, 
\end{align}
where $E=-p_\mu \xi_0^\mu$ and $L_{i}=p_\mu \xi^\mu_{i} $ denote the energy and the angular momenta of the geodesics, and 
$\mu (\ge 0)$ is the particle mass, respectively. 
Inserting into the Hamilton-Jacobi equation, we get the separable equations
\begin{align}
\left(\frac{\D \ma S_\theta }{\D \theta }\right)^2=&\, C-\frac{(L_{3}+2n \beta \cos\theta E)^2}{\sin^2\theta }\,, \label{eq:separable_theta}\\
\left(\frac{\D \ma S_r }{\D r}\right)^2=&\, -\frac{C}{r^2+m_0^2}+\frac{E^2}{f(r)^2}-\frac{\mu^2}{f(r)}\,.
\label{eq:separable_r}
\end{align}
Here, $C(\ge 0) $ represents the separation constant related to the Casimir of ${\rm SU}(2)$. 
Equivalently, the Killing tensor $K_{\mu\nu}$ satisfying $K^{\mu\nu}p_\mu p_\nu =C$  and $\nabla_{(\mu}K_{\nu\rho)}=0$ 
is reducible and expressed in terms of symmetric product of Killing vectors (\ref{KVs}) as 
$K^{\mu\nu}=\sum_{i=1}^3 \xi_{i}^\mu \xi_{i}^\nu-4n^2\beta^2 \xi_{0}^\mu\xi_{0}^\nu$, so that we have $C=\sum_{i=1}^{3}L_{i}^{2}-4n^{2}\beta^{2}E^{2}$.

\begin{figure}[t]
\begin{center}
\includegraphics[width=7cm]{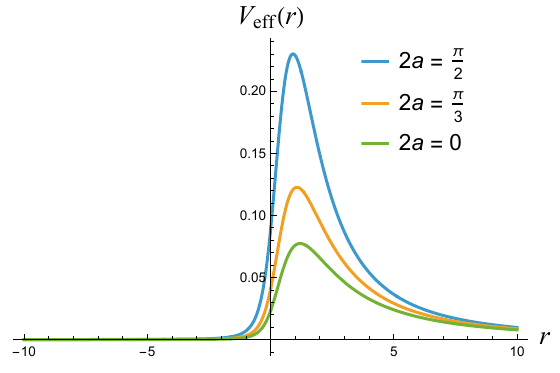}
\caption{Typical behavior of effective potential for null geodesics. The plot is shown for $m_0=1$, $\beta=0.6$, $C=1$, and $\mu=0$, with three values of the rotation angle (\ref{Omat}), $2a=\pi/2$ ($m=0, n=1$), $\pi/3$ ($m=1/2, n=\sqrt 3/2$), and $0$ ($m=1, n=0$).}
\label{fig:Veff}
\end{center}
\end{figure}

In particular, the radial equation is reduced to 
\begin{align}
\label{dotr}
\dot r^2 =E^2-V_{\rm eff}(r) \,,  
\end{align}
where the dot denotes the differentiation with respect to $\lambda$, and 
the effective potential $V_{\rm eff}(r) $ takes the form
\begin{align}
\label{effpotential}
V_{\rm eff}(r) \equiv \mu^2 f(r)+\frac{C}{r^2+m_0^2}f(r)^2 \,. 
\end{align}
Figure \ref{fig:Veff} shows the typical behavior of the effective potential $V_{\rm eff}(r)$.
One finds that $V_{\rm eff}(r)$ is bounded and regular for all $r\in \mathbb R$, 
and asymptotes to $V_{\rm eff}(r)\to \mu^2$ ($r\to \infty$) and 
 $V_{\rm eff}(r)\to \mu^2\kappa $ ($r\to -\infty$). 
It follows that the asymptotic regions $r\to \pm \infty$ can be reached 
only with an infinite affine parameter. Since $f(r)$ is smooth and strictly positive, 
the effective potential $V_{\rm eff}(r)$ is regular and bounded for finite $r$, 
while reaching either asymptotic end requires an infinite affine parameter. 
Hence, no causal geodesic terminates at a finite affine parameter and the 
spacetime is geodesically complete.

For the null geodesics with $C\ne 0$, the effective potential admits a unique maximal value and 
its locus is determined by 
\begin{align}
\label{pseq}
 e^{4\beta U(r)}(m_0+m)(r-2m_0 \beta )+(m_0-m)(r+2m_0 \beta )=0\,. 
\end{align}
This corresponds to the unstable circular photon orbit and defines the photon surface. 
By the same argument carried out in section~\ref{sec:energycond}, 
there exists a unique solution $r=r_{\rm ph}$ to this equation in the range $-2m_0\beta <r_{\rm ph}<2m_0\beta $. 
Taking the vanishing NUT parameter $n\to 0$, one recovers $r_{\rm ph}=2 m_0\beta$, 
corresponding to twice the throat coordinate $r_{\rm th}=m_0\beta$. 
It is worth mentioning that the effective potential (\ref{effpotential}) does not allow a bound state, 
which distinguishes the present solution from the overcharged Reissner-Nordstr\"om-Taub-NUT wormhole~\cite{Clement:2015aka}.

It is instructive to compare the radial locations of the throat and the photon surface.
For $\beta\ge0$, equations (\ref{throateq}) and (\ref{pseq}) imply that
\begin{align}
r_{\rm ph}<r_{\rm th}<0\,,
\end{align}
if and only if
\begin{align}
\label{betac}
0<\beta<\beta_{\rm c}\,, \qquad
\beta_{\rm c}\equiv
\frac{1}{2\pi}\ln\left(\frac{m_0-m}{m_0+m}\right)\,.
\end{align}
This interval is nonempty only for $m<0$. At either endpoint of this interval, $\beta=0$ or $\beta=\beta_{\rm c}$,
the photon surface coincides with the throat at $r=0$.
For a wormhole, it should be noted that whether the photon surface is regarded as lying
``inside'' or ``outside'' the throat depends on which asymptotic end is taken as the reference. 
In the parameter range (\ref{betac}), the photon surface lies inside the throat 
relative to the $r\to+\infty$ end, but outside it relative to the $r\to-\infty$ end.

In figure~\ref{fig:Veff}, we can see that the maximum value of the effective potential for null geodesics increases if we increase the NUT parameter $n$ for fixed wormhole mass parameter $m_{0}$ and particle's conserved quantity $C$. 
Equations~(\ref{eq:separable_theta}) and (\ref{eq:separable_r}) imply that there is a null circular geodesic restricted on the cone $\theta=$constant with $L_{1}=L_{2}=0$. 
We can identify this geodesic with a photon ring, for which its impact parameter $b=L_{3}/E$ and azimuthal angle $\theta $ are given by $b=\pm b_{\rm ph}$ and $\theta=  \theta_{\rm ph}, \pi-\theta_{\rm ph}$ with
\begin{eqnarray}
b_{\rm ph}\equiv \sqrt{4n^{2}\beta^{2}+
\frac{r_{\rm ph}^2+m_{0}^{2}}{f(r_{\rm ph})^{2}}}, \quad 
\theta_{\rm ph}= \cos^{-1}\left(- \frac{2n\beta}{b_{\rm ph}}\right), 
\end{eqnarray}
respectively (cf. Ref.~\cite{Halla:2020hee}).
Figure~\ref{fig:Veff} suggests that 
there is a unique minimum in the optical size of the photon ring 
as we vary the NUT parameter $n$ if we can interpret the impact parameter as we do in asymptotically flat spacetimes. Investigating null and timelike geodesics in this spacetime in further detail will be very interesting.

Let us next consider the null geodesics with $\mu=C=0$, for which 
$\theta ={\rm constant}$ along the geodesics. 
In this case, $\dot r=\pm E$ and $L_{3}=-2n\beta\cos\theta E$. 
The corresponding null vectors are 
\begin{align}
\label{}
p_\mu ^{(\pm)}\D x^\mu =-E (\D t+2n \beta \cos\theta \D \phi)\pm \frac{E}{f}\D r \,. 
\end{align}
These vectors are shear-free and correspond to the principal null congruences 
for the Weyl tensor $p^{(\pm)\nu}p^{(\pm )\rho }C_{\mu\nu\rho[\sigma}p^{(\pm)}_{\tau]}=0$ \cite{Wald:1984rg}.

A notable difference between the present spacetime and the ordinary Taub-NUT 
can be understood directly from the geodesic equations. Suppose counterfactually 
that $f(r)$ possessed a simple zero at $r=r_h$ where $f(r) \sim f'(r_h)(r-r_h)$.  
From (\ref{dotr}) we have $\dot r\sim \pm E $ around $r\sim r_h$.  
On the other hand,  
\begin{align}
\label{}
\dot t=\frac{E}{f(r)} -2n\beta \cos\theta \dot \phi \,, 
\qquad 
\dot \phi=\frac{f(r)}{r^2+m_0^2}\frac{L_{3}+2n\beta E \cos\theta }{\sin^2\theta }\,, 
\end{align}
imply $\dot \phi \sim 0$ and $\dot t \sim E/[f'(r_h)(r-r_h)]$. 
It follows that $t$ diverges as $t\sim \pm \ln |r-r_h|$ as $r\to r_h$. 
This logarithmic divergence is similar to that encountered in the Schwarzschild black hole, 
simply implying that ($t, r$) coordinates do not cover the entire spacetime. 
A notable difference, however, arises once the time coordinate is periodically identified. 
Under this identification, the divergence corresponds to an infinite winding around the timelike fibre direction within a finite affine parameter. 
This is the familiar behavior encountered near the Killing horizons of the ordinary Taub-NUT spacetime and leads either to two inequivalent Hausdorff analytic extensions, each of which is geodesically incomplete~\cite{Hawking:1973uf}, or to a geodesically complete but non-Hausdorff extension~\cite{Miller:1971em}.

In the present spacetime (\ref{metric}), however, $f(r)$ is strictly positive for 
all $r\in \mathbb R$. Therefore, the NUT charged wormhole does not allow the horizon-induced infinite windings. 
In particular, there exist no fixed points under periodic identification $t \to t+P$.
Hence, no additional fixed-point singularities arise from the periodic identification.

\section{Asymptotically locally AdS solutions}
\label{sec:AdS}

We next explore the corresponding solutions with the asymptotically locally AdS boundary conditions. 
As a starting point, it is useful to recall the Taub-NUT-AdS solution of
Einstein gravity with a negative cosmological constant
\begin{align}
\label{TNAdS}
\D s^2=-f_\Lambda (r)(\D t+2 n \cos\theta \D \phi)^2+\frac{\D r^2}{f_\Lambda (r)}+(r^2+n^2)
(\D \theta^2+\sin^2\theta \D \phi^2)\,, 
\end{align}
where 
\begin{align}
\label{}
 f_\Lambda (r)=\frac{r^2-2m r-n^2}{r^2+n^2}+g^2\frac{r^4+6n^2r^2-3n^4}{r^2+n^2}\,. 
\end{align}
This metric is asymptotically locally AdS and satisfies $R_{\mu\nu}=-3g^2 g_{\mu\nu}$, 
where $g$ is the inverse AdS radius. The solution shares many characteristic properties 
with its asymptotically flat counterpart, including the presence of a Dirac-Misner string, infinite winding of geodesics and so on. 

To seek an AdS counterpart of the NUT-charged wormhole while retaining
a nontrivial phantom scalar field, it is natural in the context of moduli fields in supergravity to introduce a scalar
potential rather than merely adding a negative cosmological constant.
The latter possibility has been explored in \cite{Blazquez-Salcedo:2020nsa,Wu:2022gpm}, where asymptotically AdS
wormhole solutions were obtained in the presence of a cosmological constant. 
Here, we  consider
\begin{align}
\label{LagAdS}
\ma L= R+  \frac 12 (\nabla \varphi)^2 -2V(\varphi )\,,
\end{align}
where 
an AdS vacuum  $\Lambda=V(\varphi_*)<0$ is realized 
at the critical point $\partial V/\partial \varphi|_{\varphi=\varphi_*}=0$ of the potential.

Although asymptotically AdS spacetimes play a central role in holography,
the construction of the asymptotically AdS exact solutions is a formidable task in general. 
In particular, the scalar potential, or even a cosmological constant, partially or completely
breaks the coset symmetry underlying the solution-generating technique
employed in section~\ref{sec:asyflat}. Consequently, this technique is no longer 
applicable in the present setting~\cite{Klemm:2015uba}.

\subsection{Construction procedure for static solutions}

In spite of the adversity described above, certain static asymptotically AdS solutions can be
constructed from their asymptotically flat counterparts. The method
developed in \cite{Nozawa:2020gzz} first rewrites the asymptotically flat
seed solution in isotropic coordinates, embeds it into a McVittie-type
asymptotically de Sitter ansatz, and then determines the scalar potential
from the field equations. Going back to the static coordinates, a subsequent analytic continuation of the
Hubble parameter yields the corresponding asymptotically AdS solution. 
Depending on the sign of the scalar field, 
this prescription gives rise to two distinct branches  of asymptotically AdS Ellis-Bronnikov wormholes\footnote{The origin of the two branches in (\ref{EBAdSmetric}) with (\ref{Deltapm0}) is not immediately apparent from the static construction itself. 
It can, however, be understood from the accelerating wormhole solution found in~\cite{Nozawa:2023aep}. 
This solution belongs to the C-metric class and describes accelerating wormholes. In the
zero acceleration limit, it reduces to one of the two branches in (\ref{Deltapm0}).
The C-metric form also admits a discrete transformation that interchanges
the coordinates $x$ and $y$ accompanied by a double Wick rotation
of the two Killing coordinates. Applying this transformation before taking the zero-acceleration limit yields the other branch in
(\ref{Deltapm0}). Thus, the two static branches can be regarded as two inequivalent zero-acceleration limits related by the flipping symmetry
of the parent C-metric.} as \cite{Nozawa:2020gzz} (see also \cite{Huang:2020qmn})
\begin{align}
\label{EBAdSmetric}
\D s^2=&\, -f_0(r)\Delta_\pm^{(0)}  (r) \D t^2+\frac 1{f_0(r)} \left[
\frac{\D r^2}{\Delta_\pm^{(0)}  (r)}+(r^2+m_0^2) (\D \theta^2+\sin^2\theta \D \phi^2)
\right]\,, \\
\varphi=&\, \pm 2\sqrt{1+\beta^2}U(r) \,, 
\end{align}
where $f_0$ and $U$ are given by (\ref{EBmetric}) and (\ref{U}) respectively, and 
\begin{align}
\label{Deltapm0}
\Delta_+^{(0)}(r)=1+g^2 (r^2+m_0^2)e^{4\beta U(r)} \,, \qquad 
\Delta_-^{(0)}(r)=1+g^2 \left[(r+2m_0 \beta )^2+(1+4\beta ^2)m_0^2\right]\,. 
\end{align}
Both branches of solutions are supported by the scalar potential 
\begin{align}
\label{V}
V(\varphi)=4 \left(-4 \left(\frac{\partial W}{\partial \varphi}\right)^2-3 W^2 \right)\,, 
\end{align}
where $W(\varphi)$ plays the role of superpotential as in the supergravity context, and is given by 
\begin{align}
\label{W}
W(\varphi)=\frac g2 e^{\frac{\beta \varphi}{2\sqrt{1+\beta^2}}}\left[
\cos\left(\frac{\varphi}{2\sqrt{1+\beta^2}} \right)-\beta\sin \left(\frac{\varphi}{2\sqrt{1+\beta^2}} \right)
\right] \,. 
\end{align}
Here, $g(>0)$ is a constant corresponding to the inverse of the AdS radius at the origin $\varphi=0$. 
This (super)potential is obtained by a suitable analytic continuation of the ${\cal N}=2$
gauged supergravity model \cite{Nozawa:2020gzz}. See figure 2 in~\cite{Nozawa:2023aep} for the shapes of the functions $V(\varphi)$ and $W(\varphi)$.

One finds that 
the potential $V(\varphi)$ admits an infinite number of AdS critical points  $\varphi_p $ and $\ti \varphi_p$  ($p\in \mathbb Z$) as 
\begin{align}
\label{cpts}
\varphi_p= 2\sqrt{1+\beta^2}p \pi \,, \qquad 
\ti \varphi_p= \left\{
  \begin{aligned}
  & 2\sqrt{1+\beta^2}\left[p \pi+\arctan\left(\frac 1{2\beta }\right)\right]
 \qquad &(\beta\ne 0)\,,
  \\
  & (2p+1)\pi 
  \qquad &(\beta= 0)\,.
  \end{aligned}
\right.  
\end{align}
The local minima $\varphi=\varphi_p$ of the potential correspond also to the critical points of the superpotential 
$\partial W/\partial \varphi|_{\varphi=\varphi_p}=0$. By contrast,  the local maxima $\ti \varphi_p$ do not 
extremize the superpotential. 
At $\varphi=\varphi_p$, we find 
\begin{align}
\label{mass}
V(\varphi_p)=-3 g_p^2 \,, \qquad g_p \equiv g e^{p\pi \beta} \,, \qquad \mathsf m_p^2 \equiv -2 V''(\varphi_p)=-2 g_p^2 \,.
\end{align}
Thus, the AdS radius associated with the vacuum $\varphi_p$ is $g_p^{-1}$. 
The wormhole solution (\ref{EBAdSmetric}) is asymptotically AdS at both ends $r\to \pm \infty$, where 
the $r\to \infty$ case corresponds to $\varphi=0$, whereas the 
$r\to -\infty $ case corresponds to different vacua $\varphi=\varphi_{\pm 1}$~\cite{Nozawa:2023aep}.

An important and noteworthy feature of this construction is that the parameter $\beta$,  which is an integration constant characterizing the asymptotically flat Ellis-Bronnikov solution (\ref{EBmetric}), is promoted to a parameter
of the theory through the scalar potential (\ref{V}). Consequently, once the
potential is fixed, $\beta$ is no longer a freely adjustable parameter of the asymptotically AdS solution.
Since this identification is a genuine restriction, the resulting asymptotically AdS solutions 
should not be regarded as the most general family of corresponding asymptotically AdS solutions.
This is the price to pay to construct an asymptotically AdS solution within the present construction.

\subsection{Asymptotically locally AdS wormholes with NUT charge}

Motivated by the structure of the static solutions above, we now
extend the construction to the stationary case. In the static
solutions, the effect of the scalar potential is encoded entirely in
the ``blackening factor'' $\Delta^{(0)}_\pm(r)$ in (\ref{EBAdSmetric}) \cite{Nozawa:2020gzz}, while the remaining
functions retain their asymptotically flat form (\ref{EBmetric}).
We therefore consider the ansatz
\begin{align}
\label{metricAdS}
\D s^2=&\,-f (r) \Delta_\sigma (r) (\D t+2n\beta \cos \theta \D \phi )^2 
+\frac 1{f(r)}\left[\frac{\D r^2}{\Delta_\sigma (r) }+(r^2+m_0^2)(\D \theta ^2 +\sin^2 \theta \D \phi^2)\right]\,, 
\\
\label{scalarAdS}
\varphi=&\,  2\sigma \sqrt{1+\beta ^2 }U(r)\,,
\end{align}
where $U$ and $f$ are given by (\ref{U}) and (\ref{fpsi}), and the parameters $m$, $n$ satisfying the constraint (\ref{constraint}). 
Namely, these quantities remain the same as in the asymptotically flat case. 
The  influence of the scalar potential manifests itself only in the factor $\Delta_\sigma (r)$.
Here, we have introduced a notation $\sigma=\pm 1$ indicating two branches. 
This prescription indeed works for a hairy Taub-NUT solution in gauged supergravity \cite{Nozawa:2026nod}.
Substituting the above configurations into field equations derived from the Lagrangian (\ref{LagAdS})
with a potential (\ref{V}) and (\ref{W}), we find 
\begin{align}
\label{}
\Delta_\sigma (r)=1+g^2 \left\{\frac{(m_0+\sigma m)(r^2+m_0^2)e^{4\sigma \beta U(r)}}{2m_0}+\frac{(m_0-\sigma m)[(r- 2\sigma m_0 \beta )^2+(1+4\beta^2)m_0^2]}{2m_0}\right\}\,.
\end{align}
In the vanishing NUT limit $n\to 0$, we have $m_0\to|m|$.
Choosing $m>0$, so that $m_0\to m$, the two functions
$\Delta_\sigma(r)$ ($\sigma=\pm 1$) reduce respectively to the two static branches
$\Delta^{(0)}_\pm (r)$ given in (\ref{Deltapm0}). 

On account of $m_0\pm \sigma m\ge 0$, we find
$\Delta_\sigma (r)$ is positive-definite in both branches, thereby 
the stationary Killing field $\partial/\partial t$ is globally timelike for $r\in \mathbb R$. 
There appear neither curvature singularities nor infinite windings even after the periodic identification of time coordinate, but the 
CTCs are ubiquitous. Thus, the solution can be identified as a nutty wormhole 
in the asymptotically locally AdS spacetime.

\subsubsection{Asymptotic region $r\to \infty$}

In terms of the areal radius $S=\sqrt{g_{\theta\theta}}$, the metric and the scalar field are expanded asymptotically 
$r\to \infty$ as 
\begin{align}
\label{}
\D s^2\simeq &\, -\left(1+4g^2 n^2\beta^2-\frac{2\mathsf M_{+}^{(\sigma)}}{S}+g^2 S^2\right)(\D t+2n\beta \cos\theta \D \phi)^2
\notag \\
& + \frac{\D S^2}{1+\gamma_{+}^{(\sigma)}+g^2 S^2 -2\mathsf M_{+}^{\prime(\sigma)}/S}+S^2 (\D \theta^2+\sin^2\theta \D \phi^2) \,, 
\label{asyexmetric1}
\\
\varphi\simeq &\, \frac{X_{+}^{(\sigma)}}{S}+\frac{Y_{+}^{(\sigma)}}{S^2} \,, \label{asyexscalar1}
\end{align}
where $X_{+} ^{(\sigma)}$ and $Y_{+}^{(\sigma)}$ denote the coefficients of the $S^{-1}$ and $S^{-2}$ scalar modes, 
and are given by 
\begin{align}
\label{}
X_{+} ^{(\sigma)}=  2\sigma\sqrt{1+\beta^2}m_0 \,, \qquad 
Y_{+}^{(\sigma)}= m\beta X_{+} ^{(\sigma)} \,. 
\end{align}
Other parameters are found to be 
\begin{align}
\label{Mp}
\mathsf M_{+} ^{(\sigma)}\equiv &\,m\beta -\frac 23 g^2 \beta \left[6m^3\beta^2-mm_0^2(1+10\beta^2)+\sigma m_0^3(1+4\beta^2)\right]\,, 
\\
\label{Mpp}
\mathsf M_{+} ^{\prime(\sigma)} \equiv &\,\mathsf M_{+} ^{(\sigma)}+\frac 13 g^2 X_{+} ^{(\sigma)}Y_{+} ^{(\sigma)}\,,
\\
\gamma_{+}^{(\sigma)} \equiv &\, 3g^2 n^2\beta^2-\frac 14 g^2 (X_{+} ^{(\sigma)})^2\,.
\label{gammap}
\end{align} 
Note that the unfamiliar terms proportional to $g^2 n^2$ in $g_{tt}$ and $g_{SS}$ are 
also present in the expansion of ordinary Taub-NUT-AdS solution (\ref{TNAdS}).

The scalar field approaches the AdS critical point $\varphi=\varphi_0(=0)$. 
The spacetime, however, is only asymptotically locally AdS, since the NUT fibration persists at the conformal boundary.
Indeed, it follows from (\ref{asyexmetric1}) that the boundary metric is not a cylinder $\mathbb R\times S^2$ but deformed to be 
\begin{align}
\label{}
\D \bar s^2 _3 =-(\D t+2n\beta \cos\theta \D \phi)^2+g^{-2}(\D \theta^2+\sin^2\theta \D \phi^2)\,. 
\end{align}
For fixed theory parameters $g$ and $\beta$, the preservation of this boundary metric requires the NUT parameter $n$ to be fixed. 
Observe that the conformal boundary itself inherits the causal pathology associated with the NUT fibration. 

Let us next explore the conserved mass of the spacetime. To this end, the main difficulty is that 
the scalar field is slowly decaying at infinity and both asymptotic modes in (\ref{asyexscalar1}) are normalizable, 
since the mass parameter (\ref{mass}) of the scalar field lies in a window $-\frac 94 g^2 <\mathsf m^2_p <-\frac 54 g^2$ 
\cite{Breitenlohner:1982jf,Ishibashi:2004wx}. 
Consequently, the conserved charge depends not only on the metric coefficients but 
also on the boundary condition relating $X_+^{(\sigma)}$ and $Y_+^{(\sigma)}$. 
For $n=0$, the spacetime satisfies the relaxed asymptotically AdS boundary conditions
of \cite{Hertog:2004dr,Henneaux:2006hk}, rather than the standard Henneaux-Teitelboim
fall-off conditions~\cite{Henneaux:1985tv}.
These relaxed conditions accommodate the scalar backreaction while preserving
the full AdS asymptotic symmetry group ${\rm SO}(3,2)$.
The mass defined in this framework is $\mathsf M_+^{(\sigma)}$.

Here, we evaluate the conserved charge associated with the stationary Killing field  $\xi =\partial/\partial t$
 using the covariant phase space method~\cite{Iyer:1994ys}. 
The variation of the Hamiltonian $H_\xi$ is given by 
\begin{align}
\label{deltaH}
\delta H_\xi=\int_{S^\infty} \delta \boldsymbol Q-i_\xi \boldsymbol \theta\,,
\end{align}
where $i_\xi $ is an interior product, contracting the first index with $\xi^\mu$ and 
$\boldsymbol Q$ represents the Noether charge 2-form 
\begin{align}
\label{}
\boldsymbol Q_{\mu\nu}= &\,-\frac 1{16\pi} \epsilon_{\mu\nu\rho\sigma}\nabla^\rho \xi^\sigma \,,
\end{align}
and 
$\boldsymbol \theta =\boldsymbol \theta^{(\rm grav)}+\boldsymbol \theta^{(\varphi)}$ is the symplectic potential 
\begin{align}
\boldsymbol \theta _{\mu\nu\rho}^{(\rm grav)}=&\,\frac 1{16\pi} \epsilon_{\sigma\mu\nu\rho}g^{\sigma\tau}g^{\kappa\lambda}
(\nabla_\kappa \delta g_{\tau\lambda}-\nabla_\tau \delta g_{\kappa\lambda}) \,,
\\
\boldsymbol \theta _{\mu\nu\rho}^{(\rm \varphi)}=&\,\frac 1{16\pi} \epsilon_{\sigma\mu\nu\rho}\delta\varphi \nabla^\sigma \varphi \,. 
\end{align}
Substituting the asymptotic expansion (\ref{asyexmetric1}) and (\ref{asyexscalar1}), 
using the relation (\ref{gammap}), it turns out that the surface integral is indeed finite, giving 
\begin{align}
\label{deltaHsol}
\delta H_\xi^{(\sigma)} =\delta \mathsf M_+^{\prime(\sigma)}-\frac 14 g^2 \left(2Y_{+}^{(\sigma)}\delta X_{+}^{(\sigma)}+X_{+}^{(\sigma)}\delta Y_{+}^{(\sigma)}\right)
=\delta \mathsf M_+^{(\sigma)}-\frac 1{12}g^2\left(2Y_{+}^{(\sigma)}\delta X_{+}^{(\sigma)}-X_{+}^{(\sigma)}\delta Y_{+}^{(\sigma)}\right)\,,
\end{align}
where we have used the relation (\ref{Mpp}) at the second equality. 
It is worth commenting that this agrees with the general expression of mass variation 
in the spherically symmetric case \cite{Lu:2013ura,Liu:2013gja}, 
if we flip the sign of the second term originating from the phantom nature.
This variation  is integrable and yields 
\begin{align}
\label{}
H_\xi^{(\sigma)}  =\mathsf M_+^{(\sigma)}  +\frac 13 g^2  \beta(1+\beta^2) mn^2\,,
\end{align}
where we have set the additive constant depending only on the fixed boundary data
$(g, \beta , n)$ to vanish.
 It can be verified that this expression agrees with the holographic computation 
of \cite{Anabalon:2015xvl}, based on the renormalized Brown-York stress tensor 
\cite{Brown:1992br} supplemented by appropriate gravitational and scalar counterterms.

When the NUT parameter vanishes $n=0$, the scalar coefficients obey $Y_{+}^{(\sigma)}\propto (X_{+}^{(\sigma)})^2$. 
This means $2Y_{+}^{(\sigma)}\delta X_{+}^{(\sigma)}-X_{+}^{(\sigma)}\delta Y_{+}^{(\sigma)}=0$, 
implying $H_\xi^{(\sigma)} =\mathsf M_+^{(\sigma)}$. In this case, the boundary conformal symmetry ${\rm SO}(3,2)$ is preserved~\cite{Hertog:2004dr,Henneaux:2006hk}. 
For $n\ne 0$, on the other hand, this homogeneous relation is lost because of $X_+^{(\sigma)}\propto \sqrt{m^2+n^2}$, 
leaving a finite scalar contribution to the Hamiltonian charge. 
Because both scalar modes are normalizable, 
the boundary condition is specified by a functional relation between
$X_{+}^{(\sigma)}$ and $Y_{+}^{(\sigma)}$, and the Hamiltonian charge generally depends on this relation \cite{Hertog:2004ns}. 
Such mixed boundary conditions generally preserve only the isometries of the chosen boundary metric, rather than the full asymptotic AdS symmetry group. Namely,  the conformal symmetry of the boundary is recovered only for special choices of the functional relation~\cite{Anabalon:2015xvl}.

Unlike in the ALF case, there is no universally accepted 
prescription for defining the NUT charge in asymptotically locally AdS spacetimes, 
where several inequivalent notions of gravitational magnetic charge have been 
proposed. Here we adopt the geometrical definition based on the normalized 
Killing one-form, following \cite{Hunter:1998qe,Wu:2023woq}, 
which amounts to (\ref{NUTchern}). Thus, we have $N_+=n\beta$. 

For black-hole spacetimes, the consistency of such definitions can be tested, 
for instance, through the first law of black-hole thermodynamics. It would be 
intriguing to identify an analogous consistency relation for wormhole geometries.

\subsubsection{Asymptotic region $r\to -\infty$}

One can similarly get the asymptotic expansion for $r\to -\infty$ as 
 \begin{align}
\label{}
\D s^2\simeq &\, -\left(1+4g^2_{\sigma} \ti n^2\beta^2-\frac{2\mathsf M_{-}^{(\sigma)}}{S}+g_{\sigma}^2 S^2\right)
\left(\D \ti t+2\ti n\beta \cos\theta \D \phi\right)^2
\notag \\
& + \frac{\D S^2}{1+\gamma_{-}^{(\sigma)}+g^2_{\sigma} S^2 -2\mathsf M_{-}^{\prime(\sigma)}/S}+S^2 (\D \theta^2+\sin^2\theta \D \phi^2) \,, 
\\
\varphi\simeq &\, \varphi_{\sigma}+\frac{X_{-}^{(\sigma)}}{S}+\frac{Y_{-}^{(\sigma)}}{S^2} \,, 
\label{asyexscalar2}
\end{align}
where $\ti t$ and $\ti n$ are  defined as before (\ref{titr}) and (\ref{tin}), with $\kappa $ given by (\ref{kappa}). 
We remind $\varphi_\sigma =2 \pi\sigma\sqrt{1+\beta^2}$ and $g_\sigma =g e^{\sigma \pi \beta}$, 
see (\ref{cpts}) and (\ref{mass}). 
Expansion coefficients are given by 
\begin{align}
\label{}
X_{ -}^{(\sigma)} \equiv -2\sigma\sqrt{1+\beta^2} \kappa^{1/2}\bar m_0\,, \qquad 
Y_{-}^{(\sigma)} \equiv -\kappa^{1/2}  \bar m \beta X_{-}^{(\sigma)} \,,
\end{align}
and 
\begin{align}
\label{}
\mathsf M_- ^{(\sigma)}\equiv &\,
-\bar m\beta\kappa^{1/2} +\frac 23 g^2_\sigma \beta \kappa^{3/2} 
\left[6\bar m^3\beta^2-\bar m \bar m_0^2(1+10\beta^2)+\sigma \bar m_0^3(1+4\beta^2)\right]\,, 
\end{align}
\begin{align}
\label{}
\mathsf M_-^{\prime(\sigma)}\equiv & \, \mathsf M_-^{(\sigma)}+\frac 13 g_{\sigma}^2 X_{-}^{(\sigma)}Y_{-}^{(\sigma)} \,,\\
\gamma_-^{(\sigma)} \equiv & \,  3 g_{\sigma}^2 \ti n^2 \beta ^2 -\frac 14 g_{\sigma}^2 (X_{-}^{(\sigma)})^2 \,.
\end{align}
To define the Hamiltonian charge at each asymptotic end, we must consider variations that preserve the conformal boundary metric at that end. 
Thus, $n$ is held fixed in the $r\to \infty$ region, whereas $\ti n$
 is held fixed in the $r\to -\infty$ region. Since $\ti n=n\sqrt{\kappa}$ 
 and $\kappa$ depends on the solution parameters, these two prescriptions do not correspond to a single variation on a common global phase space. 
 Rather, they define the conserved masses associated separately with the two asymptotic ends. 
 This is similar to what we encountered in defining the ADM masses for the Ellis-Bronnikov wormholes,
where the normalization of two time translations is different. 
 
Bearing this remark in mind, a direct computation gives  the Hamiltonian conjugate to $\partial/\partial \ti t$ as
\begin{align}
\label{Hamm}
H_{\partial/\partial \ti t}^{(\sigma)}=\mathsf M_-^{(\sigma)} -\frac 13 g_\sigma^2  \beta (1+\beta^2) \kappa^{1/2}
\bar m  \ti n^2\,. 
\end{align}

It is worthwhile to emphasize that quantities for $r\to \pm \infty$ regions are formally equivalent 
under the replacement
\begin{align}
\label{}
g\to g_\sigma\,, \qquad \sigma \to -\sigma \,, \qquad 
m\to -\kappa^{1/2} \bar m \,, \qquad 
n\to \ti n=\kappa^{1/2} n \,, \qquad 
m_0 \to \kappa^{1/2} \bar m_0=\frac{m_0}{\kappa^{1/2}}\,.
\end{align}
This provides a nontrivial consistency check on the identification of (\ref{Hamm}) as the mass associated with the
$r\to -\infty$ asymptotic end.

\section{Conclusion}
\label{sec:conc}

In this paper, we have presented a new exact solution to the Einstein-phantom-scalar system that describes a traversable wormhole 
endowed with a NUT charge. 
The NUT charge nontrivially deforms the asymptotic geometry, so that the resulting spacetime is ALF rather than asymptotically flat. 
The spacetime possesses two asymptotic regions connected by a throat in the orbit space and is stationary, horizonless, 
and free of curvature singularities. 
At the same time, it inherits characteristic global features of the ordinary Taub-NUT geometry, including the Dirac-Misner string and the presence of CTCs throughout the spacetime. 
A notable distinction from the ordinary Taub-NUT spacetime arises in the geodesic structure. 
Since the present solution possesses no Killing horizons, geodesics do not undergo the infinite winding associated with the approach to the Taub-NUT horizons even when the time coordinate is periodically identified. 
Together with the regularity of the geometry, this allows all causal geodesics to be extended to arbitrary values of their affine parameters, rendering the spacetime geodesically complete. In an appendix, we have also obtained a Euclidean continuation whose conformal class admits an ambi-K\"ahler structure.

We have also constructed an asymptotically locally AdS generalization of the NUT-charged wormhole by introducing a nontrivial potential (\ref{V}) for the phantom scalar field. 
The potential admits an infinite sequence of AdS vacua, and the resulting wormhole generally interpolates between different vacua at its two asymptotic ends. 
Owing to the NUT charge, the spacetime does not fall into the customary asymptotically AdS boundary conditions with the full ${\rm SO}(3,2)$ asymptotic symmetry group. 
Nevertheless, the covariant phase space analysis allows us to show that the Hamiltonian charge associated with time translations remains finite,
once an appropriate functional relation between the two normalizable scalar modes is imposed. 
The resulting mass depends, in general, on this choice of scalar boundary condition, in accordance with the familiar structure in the context of designer gravity
\cite{Anabalon:2015xvl,Hertog:2004ns}.

Several directions for extending the present work remain open. 
An immediate question concerns the linear stability of the NUT-charged wormholes. 
The ordinary Ellis-Bronnikov wormhole is known to suffer from a radial instability~\cite{Gonzalez:2008wd,Torii:2013xba}, 
and it would be interesting to determine how the NUT charge modifies the unstable spectrum and whether the nontrivial NUT fibration introduces additional unstable sectors. It would also be worthwhile to seek a genuinely rotating generalization of the Ellis-Bronnikov wormhole, in the Kerr sense rather than through a NUT charge. 
Such a solution may be pursued within the framework of a Kerr-Schild transformation~\cite{Nozawa:2025yac}.

On the AdS side, a more systematic analysis of admissible scalar boundary conditions and the corresponding asymptotic symmetries may clarify the role of these wormholes in holography, in particular when the two asymptotic ends approach different AdS vacua. 
Further generalizations to include electromagnetic fields, higher dimensions, or more general scalar potentials may reveal a broader class of exact nutty wormhole geometries. 
Finally, the Euclidean continuation presented in appendix suggests a possible connection with gravitational instantons and deserves a more detailed investigation.

\section*{Acknowledgments}
TH is grateful to T.~Takeuchi for fruitful discussion.
The work of MN is partially supported by JSPS KAKENHI Grant Number JP25K07309. TH was supported by JSPS KAKENHI Grant Number JP24K07027.

\appendix

\section{Euclidean solution}
\label{sec:Euc}

Although the Lorentzian Taub-NUT spacetime (\ref{TaubNUT}) suffers from a number of unconventional 
causal and global properties, its Euclidean counterpart has played a prominent role 
in gravitational physics. 
In particular, the self-dual Euclidean Taub--NUT solution provides one of the simplest 
examples of a gravitational instanton, namely, a complete and nonsingular 
positive-definite solution of the Euclidean Einstein equations 
\cite{Hawking:1976jb,Gibbons:1979xm}. 
It is ALF and possesses a nontrivial $S^1$ fibration at infinity, 
with the fixed point of the ${\rm U}(1)$ action giving rise to the ``nut'' structure \cite{Gibbons:1979xm}. 
The Taub-NUT geometry also underlies the multi-center gravitational instantons 
of Gibbons and Hawking \cite{Gibbons:1978tef}, when the curvature is self-dual, corresponding to the 
hyper-K\"ahler manifold.  
These Euclidean solutions have provided important semiclassical saddle points in 
Euclidean quantum gravity and useful laboratories for studying topology, self-duality, 
and gravitational analogues of Yang--Mills instantons 
\cite{Hawking:1976jb,Gibbons:1976ue}.
Motivated by these developments, we now examine the Euclidean continuation of the 
NUT-charged wormhole constructed in the main text.

Performing the double Wick rotations 
$t\to i t$ and $n\to i n$ to (\ref{metric}), we can obtain a solution with Euclidean signature 
\begin{align}
\label{metricEuc}
\D s^2=&\,f (r) (\D t+2n\beta \cos \theta \D \phi )^2 
+\frac 1{f(r)}[\D r^2+(r^2+m_0^2)(\D \theta ^2 +\sin^2 \theta \D \phi^2)]\,, 
\\
\label{scalarEuc}
\varphi=&\, \pm 2\sqrt{1+\beta ^2 }U(r)\,, 
\end{align}
where 
\begin{align}
\label{}
f(r)=\frac{m_0}{m_0 \cosh (2\beta U(r))+m \sinh (2\beta U(r))}\,,\qquad 
U(r)= \frac \pi 2-\arctan \left(\frac{r}{m_0}\right)\,, 
\end{align}
as in the Lorentzian case, while the constraint should be tailored to 
\begin{align}
\label{constraintEuc}
m_0=\sqrt{m^2-n^2}\,. 
\end{align}
This solution obeys the Euclidean counterpart of field equations (\ref{Eineq4D}). 
One can set $\beta \ge 0$ and $n\ge 0$ without losing generality, and (\ref{constraintEuc}) 
requires $|m|>n$. For the regularity on the axis, 
one must choose the periodicity of $t$ coordinate to be (\ref{period}), 
rendering the solution to possess ${\rm SU}(2)\times {\rm U}(1)$ isometry.

Contrary to the Lorentzian case, the denominator of $f(r)$ is not positive definite 
over the entire parameter range. The singular surface $r=r_s$ where $f^{-1}|_{r=r_s} =0$ occurs when
\begin{align}
\label{}
e^{4\beta U(r_s)}=\frac{m-m_0}{m+m_0} \,, 
\end{align}
Since $1<e^{4\beta U(r_s)}<e^{4\beta \pi}$, the singular surface appears at finite $r_s$
whenever
\begin{align}
\label{}
 m<0\,, \qquad \frac{m_0}{|m|}<\tanh (2\pi \beta)\,.
\end{align}
It follows that the solution becomes regular and ALF at both asymptotic ends $r\to \pm \infty$, 
if the parameter $m$ lies in either of the ranges
\begin{align}
\label{}
m>|n|>0 \,, \qquad m< -|n|\cosh(2\pi \beta)\,. 
\end{align}

Define the 2-forms
\begin{align}
\label{Kahlerform}
\omega_\pm \equiv (\D t+2n\beta \cos\theta \D \phi) \we \D r\pm \frac{r^2+m_0^2}{f(r)}\sin\theta \D \theta \we \D \phi \,, \qquad 
\star \omega_\pm =\pm \omega _\pm \,, 
\end{align}
where the $\pm$ sign denotes the orientation and is irrelevant to the sign choice of the scalar field (\ref{scalarEuc}). 
Writing the almost complex structures as $(J_\pm)_\mu{}^\nu=(\omega_\pm)_{\rho\mu} g^{\rho \nu}$ satisfying 
$(J_\pm)_\mu{}^\rho(J_\pm)_\rho{}^\nu = -\delta_\mu{}^\nu$, 
one can verify that  the Nijenhuis tensors for $J_\pm$ vanish, i.e., 
both of the almost complex structures are integrable~\cite{Klemm:2013eca,Nozawa:2015qea}.

Since the metric (\ref{metricEuc}) is neither Ricci-flat nor endowed with the self-dual Weyl curvature, 
the space does not correspond to the 
hyper-K\"ahler manifold, unlike the ordinary Euclidean Taub-NUT instanton. 
Nevertheless, both Hermitian structures are locally conformally K\"ahler.
The exterior derivatives of the fundamental 2-forms (\ref{Kahlerform}) are  reduced to 
\begin{align}
\label{}
\D \omega_\pm - \theta _\pm \we \omega_\pm =0 \,, 
\end{align}
where $\theta_\pm $ are called the Lee forms~\cite{Gauntlett:2003cy} and are given by 
\begin{align}
\label{}
\theta_\pm =-2\D( \ln \Omega_\pm ) \,,
\end{align} 
with 
\begin{align}
\label{}
\Omega_\pm \equiv \left(\frac{f(Z\pm 1)}{(r^2+m_0^2)(Z\mp 1)}\right)^{1/2} 
\,, \qquad 
Z\equiv \frac{m+m_0 \tanh (\beta U)}{n} \,. 
\end{align}
It follows that the Lee forms are locally exact. 
Consequently, the conformal transformations $\hat g^\pm _{\mu\nu}=\Omega_\pm^2 g_{\mu\nu}$
lead to the K\"ahler metrics with K\"ahler forms $\hat \omega_\pm \equiv \Omega_\pm^2 \omega_\pm $~\cite{Gauntlett:2003cy}. 
The conformal class of the metric (\ref{metricEuc}) therefore admits an ambi-K\"ahler structure,
defining conformally related but oppositely oriented 4-dimensional K\"ahler metrics
\cite{Apostolov:2013oza,Keaton}.

\end{document}